# Structural Modification and Variation in the Kinetics of Isoconversional Phenomena: A Study on the Effect of Gamma Irradiation on Poly (Ethylene Oxide)

Madhumita Mukhopadhyay*[1], Mou Saha[1,2] Ruma Ray[2] and Sujata Tarafdar[1]

[1]Jadavpur University, Physics Department, Condensed Matter Physics Research Centre, Kolkata – 700032, WB, India

[2]Gurudas College, Physics Department, Kolkata – 700054, WB, India

Correspondence to: madhubanerji@gmail.com (M. Mukhopadhyay)
            **Tel:** +91 3324146666x2760, +91 9433882877
            **Fax:** +91 3324148917.

<u>E-mail Addresses:</u> madhubanerji@gmail.com, sahamou14@gmail.com,
              me.rumaray@gmai.com and sujata_tarafdar@hotmail.com.

**Abstract**

Interactions of Poly (Ethylene Oxide) [PEO] and gamma irradiation of variable doses (1-30 kGy) on the thermal, crystalline and structural properties are investigated using DSC and FTIR technique. Two states of PEO, viz. powder (P-S-series) and solution (P-L-series) are subjected to irradiation and are cast into uniform film. DSC results revealed steady increment of crystallinity upto 20 kGy for P-S-series after which amorphous region increases till 30 kGy. Conversely, P-L-series shows much enhanced crystallinity retained within low regime of 7 kGy, followed by sharp declining trend till 30 kGy. DSC is also used to determine the multiple kinetic processes in an isoconversional PEO melting using Friedman differential analysis. Gamma irradiation is found to generate newer functional groups established from FTIR study. FTIR spectra of generated peaks viz. -C=O, -C=C- exhibit scission dominated irradiation induced in presence of air. Contribution of cross-linkage which is higher in liquid irradiation is also proved from FTIR. The studied observations of DSC and FTIR are correlated with polymer microstructures. Hence, selective irradiation dose could be determined with respect to the exposed state (solid or solution) of polymer and utilized to tailor the properties of PEO.

# 1. Introduction

Polyethylene oxide (PEO) is a polymer with versatile range of applications. Mixed crystalline and amorphous phase in PEO matrix possesses constrains in property optimization. On the other hand, tailoring and optimizing the proportion of such mixed phase in polymer system influences the associated properties and many of its applications [1, 2]. One of the important applications of PEO is the active role as solid polymer electrolytes (SPE) in high energy batteries and electrochemical devices [3]. Among many SPE's till date, PEO is one of the widely studied host polymer complexed with alkali salts for use in solid state batteries viz. Li-ion batteries [4-7]. Though SPE's offers innumerable advantages towards compatibility with the applicable device, insufficient ionic conductivity limits its application [8]. Researchers have proposed the possibility of improvement in ionic conductivity though: a) lowering the glass transition temperature ($T_g$) of the polymer or b) reducing the energy barrier for ionic movement. Ionic conductivity in SPE has been found to influence by the concentration of defects and fraction of amorphous/crystalline distribution of phases [9]. Many plausible strategies are reported regarding the enhancement in ionic conduction eg. addition of plasticizers, incorporating organic solvent, controlling drying process in several medium etc. [10-12]. Among these, SPE perturbation though high energy electromagnetic radiation serve certain affirmative possibilities of temperature independent, clean and homogenous reaction initiating in a stable/sterile product. However, assistances from irradiation technique could be best utilized in an affirmative manner through selective alterations of associated properties. Among innumerable techniques, high energy irradiation only has the potential to induce energy within the experimental samples in order to generate favorable changes required as per the application. The prime novelty is imbibed in the selection of appropriate doses so as to tailor the exact magnitude of energy required for desired transformations within the polymer matrix. The interaction of high energy radiation with the polymeric matrix is found to be initiated through the formation of unstable high energy species caused due to series of excitation procedures viz. ionization, radical induced chain reaction etc. Based on usual thermodynamic rules applicable to unstable high energy intermediates, the excited polymer matrix tend to regain stability through series of intermediate reactions thereby generating newer functional groups within the polymer entity [13]. As a consequence of high energy irradiation, two competing effects of scission and cross-linking of the polymer chains are broadly classified. Predominance of either scission or cross-linking is dependent on various factors eg. Irradiation dose, environment, solvent, drying process etc. Owing to such facts, divergence in the experimental results is significant even upon using similar polymer sample and irradiation doses. Therefore, it could be justified that, since high energy irradiation results in intrinsic transformations within the polymer matrix, the correlated properties eg, conductivity, microstructure, mechanical strength, relaxations etc. are also found to be highly dependent on the processing conditions, even for the same polymer.

Research work related to high energy irradiation on polymer been previously communicated by our group [14-16]. The present research intends to study the influence of selective application of high energy gamma irradiation in two aspects: a) structural modifications influencing the properties of such developed films and b) multiple kinetic processes involved in the isoconversional phenomena of such irradiated polymer species. The properties of polymer eg. viscosity, distribution of amorphous and crystalline phases, microstructural distribution etc. are expected to alter significantly as a consequence of selective energy perturbation. Polymer species subjected to high energy irradiation is found to exhibit multiple process of macromolecular distribution, cross linking, free radical formation, carbonization and oxidation [17]. Understanding of such structural arrangements within the polymer species is studied in the present context which allows tailoring of properties in a predetermined manner.

## 2. Materials and methods

### 2.1. Sample Preparation

Pure Poly-(ethylene oxide) [PEO] from B.D.H, England (Molecular Weight. $10^5$) and methanol (99.9%) were used for sample preparation. In order to study the influence of electromagnetic radiation on the intrinsic properties of polymer matrix, gamma irradiation using $^{60}$Co source with a dose rate 6.4 kGy.h$^{-1}$ was applied in two physical states of PEO viz. liquid solution and powder form. The doses of gamma irradiation were varied in the range of 1-30 kGy. The irradiated PEO powders were cast into thin films. Preparation of the film initiated with preparation of solution using both irradiated and un-irradiated PEO powder with methanol as solvent. Concentration of PEO was maintained at 0.02 g.ml$^{-1}$ and 0.04 g.ml$^{-1}$ respectively. The polymer films were prepared by solution casting technique in a polypropylene petri dish and allowed to dry in air for 5 - 6 days followed by vacuum drying for 6 - 7 days, at a pressure of 0.1 mbar. Vacuum drying produces a self –supporting film, eliminating any moisture which may have been absorbed by the sample. The details of film samples prepared along with their ID's are given in Table 1.

### 2.2. Characterizations of PEO film prepared from irradiated powder and solution

The experimental samples viz. PEO films prepared from irradiated powder and solution along with the initial unirradiated form were subjected to X-ray diffraction [Bruker-AXS type diffractometer with Cu-K$\alpha$] with a scan rate of 2$^{o}$.min$^{-1}$. Differential scanning Calorimetry (DSC) was performed on all aforementioned experimental samples in the temperature range of - 65$^{o}$C-100$^{o}$C using [Pyris Diamond DSC, Perkin Elmer]. Two heating and cooling cycles ($\beta_1$ & $\beta_2$) were performed for each sample with a scan rate of 20$^{o}$C.min$^{-1}$.

**Table 1:**
Sample identification of experimental PEO samples subjected to gamma irradiation

| S. l. no. | Experiment condition of polymer sample | | Irradiation Dose (kGy) | Sample Identification |
|---|---|---|---|---|
| 1 | Unirradiated PEO | 0.02 g.ml$^{-1}$ | 0 | P-2Un |
| | | 0.04 g.ml$^{-1}$ | 0 | P-4Un |
| 2 | Powder Irradiation | 0.02 g.ml$^{-1}$ | 1 | P-2S1 |
| | | | 3 | P-2S3 |
| | | | 5 | P-2S5 |
| | | | 7 | P-2S7 |
| | | | 10 | P-2S10 |
| | | | 15 | P-2S15 |
| | | | 20 | P-2S20 |
| | | | 30 | P-2S30 |
| | | 0.04 g.ml$^{-1}$ | 1 | P-4S1 |
| | | | 3 | P-4S3 |
| | | | 5 | P-4S5 |
| | | | 7 | P-4S7 |
| | | | 10 | P-4S10 |
| | | | 15 | P-4S15 |
| | | | 20 | P-4S20 |
| | | | 30 | P-4S30 |
| 3 | Liquid Irradiation | 0.02 g.ml$^{-1}$ | 1 | P-2L1 |
| | | | 3 | P-2L3 |
| | | | 5 | P-2L5 |
| | | | 7 | P-2L7 |
| | | | 10 | P-2L10 |
| | | | 15 | P-2L15 |
| | | | 20 | P-2L20 |
| | | | 30 | P-2L30 |
| | | 0.04 g.ml-1 | 1 | P-4L1 |
| | | | 3 | P-4L3 |
| | | | 5 | P-4L5 |
| | | | 7 | P-4L7 |
| | | | 10 | P-4L10 |
| | | | 15 | P-4L15 |
| | | | 20 | P-4L20 |
| | | | 30 | P-4L30 |

[occasionally 5$^{o}$C.min$^{-1}$ for cooling cycle]. The prepared films from unirradiated powder and irradiated powder or liquid sample was studied for their structural analysis using Fourier

transformed- Infra red spectroscopy (FT-IR) [FTIR-8400S, Shimadzu, Tokyo, Japan]. The microstructural analyses of the prepared films were performed using Scanning electron microscopy (SEM) [configuration no. QUO-35357-0614] and polarizable microscope [LAS EZ, Leica Application suit, Version: 2.0.0, 2010]. The reported experimental results are found to be reproducible for all the doses.

## 3. Results and Discussion

### 3.1. Variation in crystallinity of PEO irradiated in powder and solution state

DSC results exhibits changes in the degree of crystallinity ($\chi_{c,H}$) of experimental PEO films subjected to gamma irradiation of variable doses as shown in Figs. 1a and b.

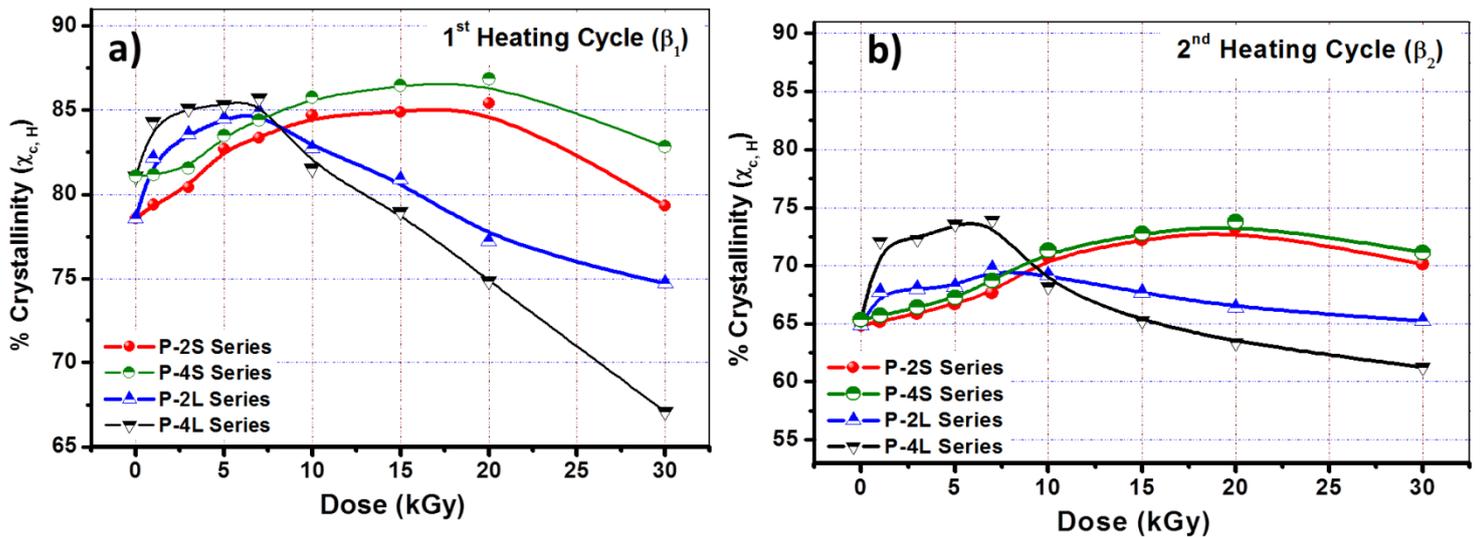

**Fig. 1.** Variation of % crystallinity obtained from DSC as a function of irradiation doses for different sample compositions for: a) first and b) second heating cycle.

The degree of crystallinity ($\chi_{c,H}$) of the experimental PEO films is calculated using the formula;

$$\chi_{c,H} = \frac{\Delta H_m}{\Delta H_{m,0}} \quad (1)$$

where, $\Delta H_m$ is the enthalpy of melting per gram of PEO and $\Delta H_{m,0}$ is the enthalpy of melting per gram of 100 % crystalline PEO ($\Delta H_{m,0}$ is used as 197 J.g$^{-1}$ [18]).

The experimental samples are subjected to two heating and cooling cycles ($\beta_1$ and $\beta_2$). The DSC outcome obtained from the first heating cycle gives information regarding the actual state of polymer crystals, whereas the subsequent cooling cycles tend to erase the previous thermal history eg. annealing during processing. The second heating and cooling cycle does not allow the semi crystalline polymer sufficient time to reorganize. This fact could be visualized from the reduced % crystallinity of the polymer films having variable history of irradiation in second heating cycle (Fig. 1b). It is observed from both Figs. 1a and b that, PEO films obtained from powder irradiated samples (P-2S and P-4S-series) shows increasing trend of % crystallinity with irradiation dose upto 20 kGy. Gamma irradiation in presence of air favors degradation of polymer fragments. The increase in crystallinity upto 20 kGy is usually explained by easier crystallization of shorter polymer chains generated from degradation of the more radiation sensitive amorphous phases in either first or second heating cycle. However, at a higher dose ($\geq$ 30 kGy), an increasing number of defects offsets the effects that have caused the increase in crystallinity resulting in the overall decrease of crystalline regions of P-2S and P-4S –series. In addition, though degradation effects seem to dominate in PEO films obatined from solid state irradiation in presence of oxygen scavenger, the simultaneous effect of cross-linkage is also evident. The influence of cross-linkage of the degraded polymer fragments tends to reduce the crystallinity at higher dose of 30 kGy. Researchers have studied the influence of tie molecules within polymer matrix towards the crystallinity or amorphosity upon irradiation [19, 20]. Tie molecules are the polymer chains in the interfacial regions that connect the crystallites. In the present case, recrystallization of such tie molecules at lower doses could be another possible factor towards the enhancement of % crystallinity upto 20 kGy. This is supported from the microstructural study in section 3.3.3. Similar trend of increasing crystallization enthalpy ($\Delta H_c$) with dose is observed upto 20 kGy followed by a reduction at 30 kGy for P-2S- and P-4S-series in both first and second heating cycle (Figs. 2a and b).

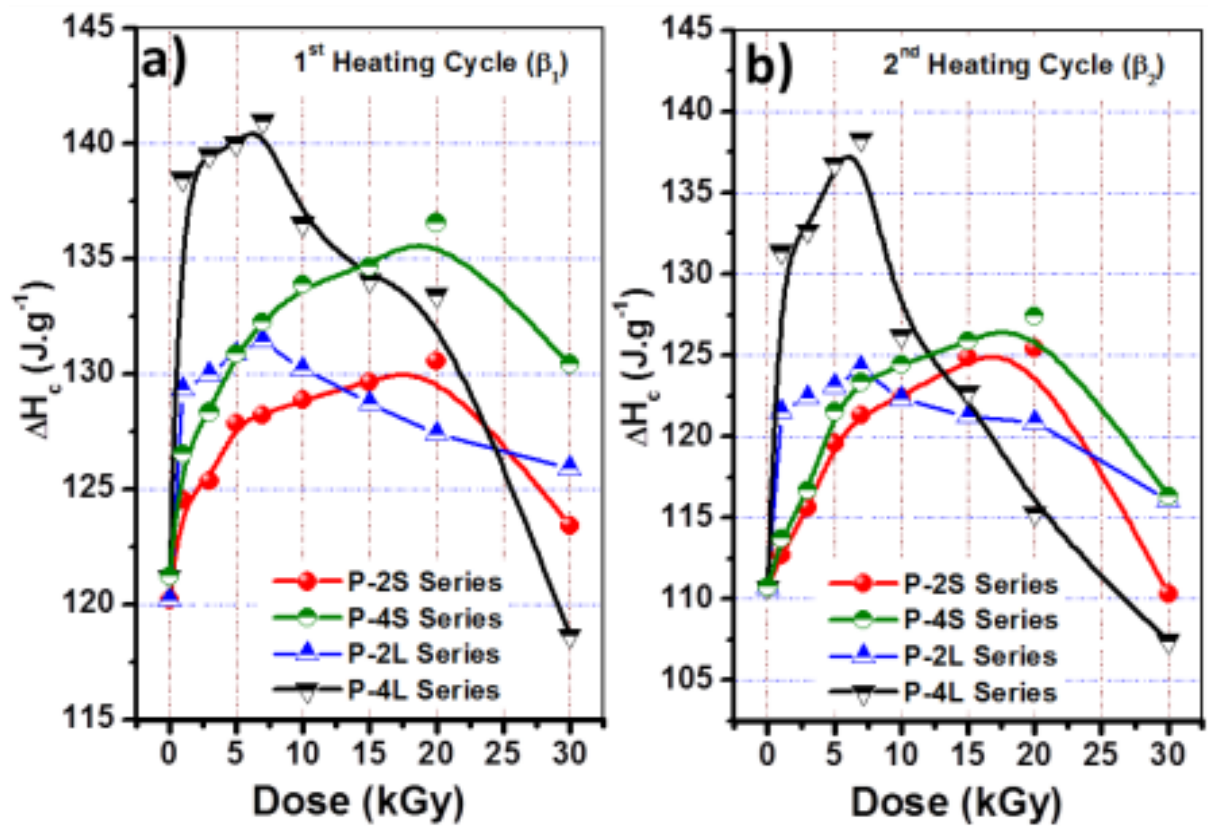

**Fig. 2.** Variation of crystallization enthalpy ($\Delta H_c$) obtained from DSC as a function of irradiation doses for different sample compositions for: a) first and b) second heating cycle.

Polymer films of P-2L and P-4L-series shows an initial increase in % crystallinity upto 7 kGy followed by a steady reduction till 30 kGy as shown in Figs. 1a and b. As discussed earlier, the magnitudes of % crystallinity is lower for second heating cycle of DSC. Reports have suggested that, polymer solutions form gel subjecting to very high gamma irradiation [21]. In the present case, the polymer solution of either P-2L or P-4L-series is found to remain liquid upto 30 kGy. However, irradiation in liquid state is expected to result in network formation, so intermolecular cross linkage is likely to be the primary radiation outcome. Air induced irradiation is found to favor degradation thereby increasing the % crystallinity initially upto 7 kGy. However, at low concentration of PEO solution (0.02 g.ml$^{-1}$), most of the radiation is absorbed by the solvent as reported by Ferloni et al [22]. In solution, irradiation is reported to generate hydroxyl radical which reacts with other formed radicals from polymer matrix. These solvent generated excited radicals tend to neutralize through cross linkage rather than β-scissions. In polymer solution of higher concentration (0.04 g.ml$^{-1}$), oxygen induced degradation is significant. In addition, solvent induced cross-linkage is found to reduce the overall crystallinity substantially beyond 10 kGy compared to P-2L-series. In solution irradiation, cross linkage is facilitated at the point of physical entanglement of polymer chains. Therefore, random distribution of polymer fragments and higher contribution of cross linkage significantly impede the crystallinity of PEO in solutions, in which large chain movements is required for crystallization (Figs. 1 a and b). Similar trend as in Fig. 1, visualized from the crystallization enthalpy as a function of dose for P-2L and P-4L-series shown in Figs. 2a and b. Considerably higher crystallization enthalpy ($\Delta H_c$) and crystallinity is observed for P-4L-series in the irradiation dose range of 1-7 kGy. Irradiation effects are more pronounced in liquid state even in lower doses due to greater mobility of PEO chains and easier diffusion of radicals. Higher concentration of polymer in mobile liquid state suffers higher degradation resulting in higher crystallization and $\Delta H_c$.

Irrespective of heating cycles, melting ($T_m$) and crystallization ($T_c$) temperatures of P-2S/4S and P-2L/4L is found to exhibit an initial increment followed by a steady decreasing trend upto 30 kGy (Figs. 3a and b). The lower magnitudes of $T_m$ and $T_c$ is greater in more amorphous samples. For P-2S and P-4S samples, in spite of the increasing trend of % crystallinity, both $T_m$ and $T_c$ decreases beyond 3 kGy steadily due to insufficient crystallization and radiation induced defects. Upto 3 kGy dose, the irradiation effect is moderate and crystallization occurs more or less uniformly which tend to increase both the melting and crystallization temperature slightly. In case of P-2L series, induced cross linkage contribution from solvent tend to restrict the molecular motion thereby decreasing $T_m$ and $T_c$ sharply at higher irradiation doses.

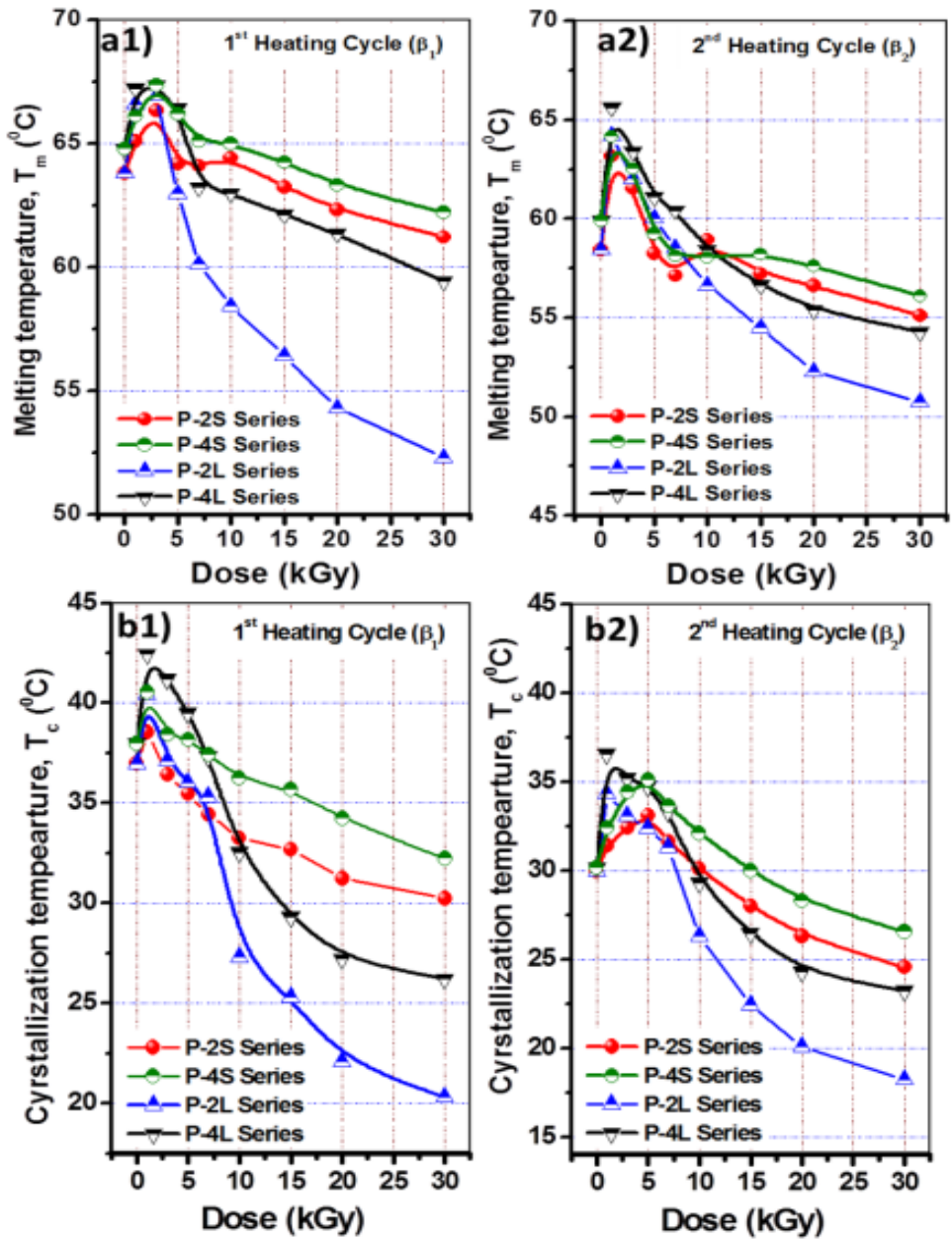

**Fig. 3.** Variation of melting ($T_m$) [a1 and a2] crystallization temperature ($T_c$) [b1 and b2] obtained from DSC as a function of irradiation doses for different sample compositions for: a1), b1) first and a2), b2) second heating cycle.

### 3.2. Study of multiple kinetic process in an isoconversional phenomena

The DSC curve shown in Fig. 4a exhibit single phase transformation of polymer during melting process termed as isoconversional phenomena. These isoconversional processes yield the values of effective activation energy as a function of conversion. A unique thermodynamic process eg. melting could be considered to comprise a number of different kinetic processes which corresponds to variable activation energy as a function of degree of conversion of the concerned process.

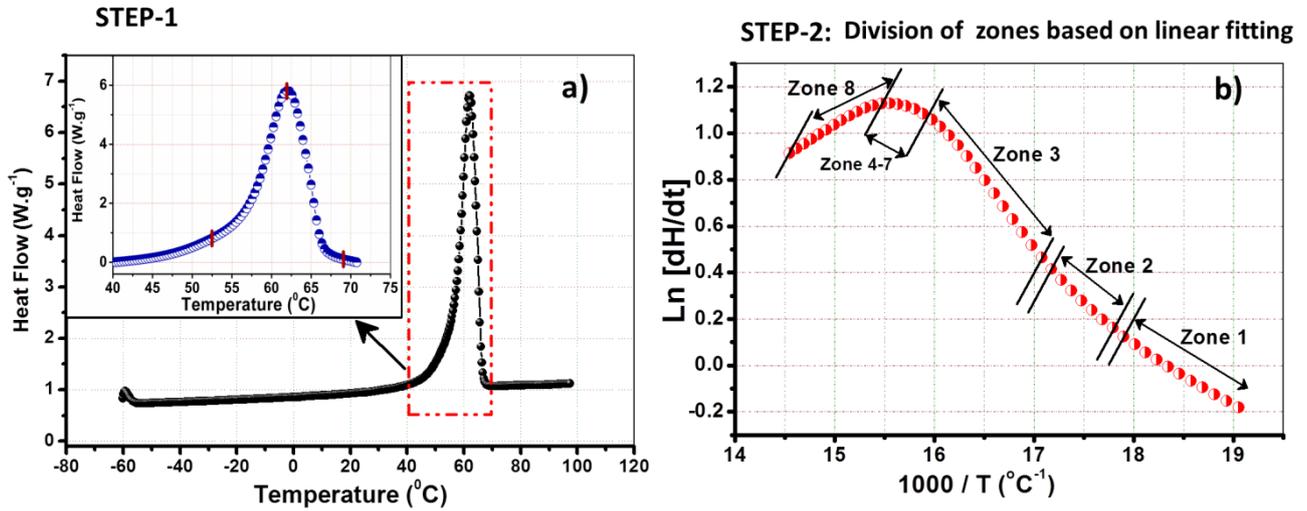

**Fig. 4.** Typical representation of: a) DSC pattern of PEO sample (INSET: selected region of the DSC curve) and b) division of zones in ln(dH/dt) vs. 1000/T plot based on linear fitting.

A complete isoconversional kinetic analysis can be carried out using various thermo analytical data eg. DSC, TGA, rheometry etc [23, 24]. Popular isoconversional techniques at different temperature programs comprises of Friedman differential method, integral method by OFW or KAS and non-linear method by Vyazovkin [25-28]. Differential method by Friedman is undertaken in the present context to elucidate the isoconversional phenomena. Thermo kinetic analysis of isoconvenrsional phenomena is performed using the equation:

$$\frac{d\alpha}{dt} = K(T)f(\alpha) \qquad (2)$$

where, $\alpha$, t, K(T) and $f(\alpha)$ are the degree of conversion, time, rate constant and kinetic model function respectively.

The dependence of rate constant [$K(T)$] is described by Arrhenius law as:

$$\frac{d\alpha}{dt} = K(T)f(\alpha) = A\exp\left(-E_a/RT\right)$$
$$\Rightarrow Ln(d\alpha/dt) = Ln\langle A_\alpha f(\alpha)\rangle - E_a/RT \qquad (3)$$

where, A, $E_a$, R and T are pre exponential factor, activation energy, gas constant and temperature.

Therefore, activation energy ($E_a$) can be estimated from the slope of Eq. 3 as a function of degree of conversion ($\alpha$). The degree of conversion ($\alpha$) is defined as the ratio between the heat exchanged at time, $t_i$ i.e $H_i$ and the total heat of concerned process (Q).

$$\alpha_i = \frac{H_i}{Q} \Rightarrow \alpha = \frac{dH}{Q} \qquad (4)$$

Eq. 3 can be written as:

$$Ln\left(dH/dt\right) = Ln\langle A_\alpha Q.f(\alpha)\rangle - \frac{E_a}{RT} \qquad (5)$$

This Friedman technique corresponds to model free approaches that allow to evaluate the Arrhenius parameters without the knowledge of any reaction model [25]. Study on the dependence of $E_a$ on the degree of conversion ($\alpha$) helps in the detection of multiple kinetic processes [25]. Fig. 4a represents a typical DSC graph of the heating cycle of PEO sample. Among the entire range temperature range of -65°C to 100°C, the exact range corresponding to the melting process is at first separated as shown in the inset of Fig. 4a. This is followed by re-plotting the selected DSC curve (inset of Fig. 4a) to the form in Eq. 5 [Ln(dH/dt) vs. 1000/ T] shown in Fig. 4b. It could be mentioned from Eq. 2 that, the determination of activation energy requires linear fitting of Ln(dH/dt) vs. 1000/ T plot. This could be achieved by dividing the obatined curve in several zones based on the linear fitting. Such fitting is performed for all the experimental samples mentioned in Table 1. Figs. 5a and b describe the variation of $E_a$ as a function of degree of conversion for the films formed from irradiated powders (P-2S and P-4S-series). Similarly, Figs. 5c and d shows the trend of $E_a$ vs. $\alpha$ for PEO films obtained from solution irradiation (P-2L and P-4L-series). Melting of PEO comprises primarily three different stages viz. ascending (Zone I1), slow ascending (Zone I2) and slow (Zone D′) and sharp descending (Zone D1 and D2). Compared to all, samples of P-2/4L5 and P-2/4L15 shows additional two stages of initial decreasing (Zone $D_0$) and flat region (Zone F). These regions signify variable kinetic stages during the melting process. Irrespective of composition, the ascending stage (Zone I1 and I2) indicates the growing contribution of higher activation energy detected by heat release in DSC.

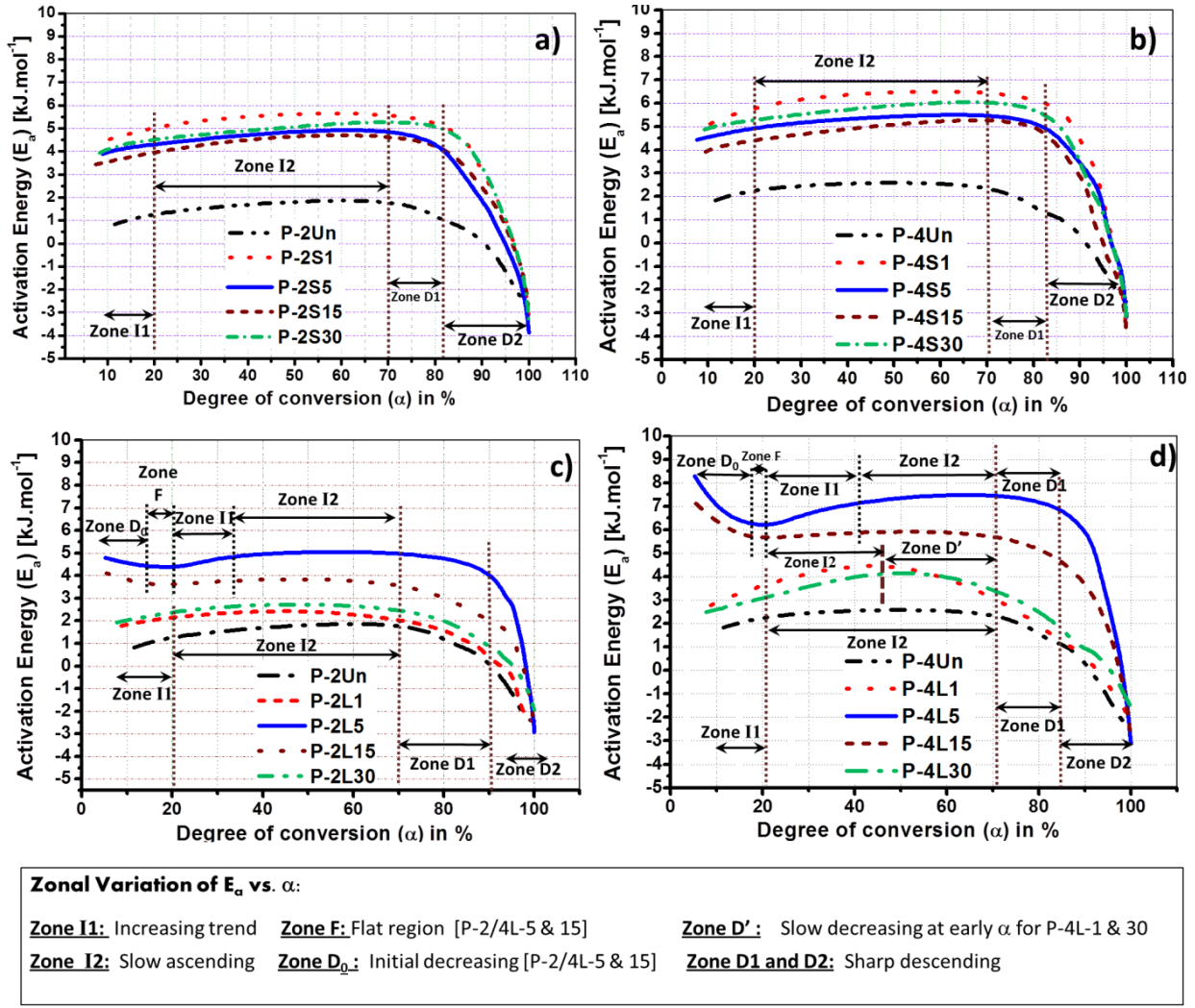

**Fig. 5.** Variation of activation energy ($E_a$) with degree of conversion ($\alpha$) as an influence of irradiation dose for: a) P-2Un and P-2S series, b) P-4Un and P-4S series, c) P-2LUn and P-2L series and d) P-4LUn and P-4L series.

As the dependence reached maxima, the plot becomes slow ascending (Zone I2) and retains from $\alpha$ = 20 % to 70 % for all samples, except, P-4L1 & 30, in which the nearly flat region ends before 45 % completion of melting process. The flat regime of $E_a$ vs. $\alpha$ dependence is followed by a descending stage in which the activation energy is much lowered. It could be observed from Figs. 5 a to d, that after $\alpha$ = 90 %, negative activation energies are obatined.

This signifies self-sustaining reaction/ process which does not require external activation, whereas, the internal energy is sufficient to complete the isoconversional process releasing certain magnitude of energy during conversion. Compared to all samples, P-4L1 and 30 shows initiation of declining pattern form $\alpha < 45\%$ and is denoted by an additional *Zone D′*. Among all the tabulated experimental films, certain sample, P-2/4L-5 & 15 shows an initial declining trend of activation upto *$\alpha < 20\%$* (Zone $D_0$) followed by a flat *Zone F* (upto *$\alpha = 20\%$*). For such samples, the polymer system initially consumes maximum energy for small $\alpha$ (observed from the highest activation energy during starting of the melting), then proceeds without barrier upto $\alpha = 20\%$. This is followed by requirement of energy before ascending of activation energy. It could be noticed from Fig. 5, that the activation requirement for P-2/4L-series is higher compared to P-2/4S-series. Trend of activation energy as a function of degree of conversion ($\alpha$) could be more clearly understood from Fig. 6 which represents ($E_a/\alpha$) [from Fig. 5] as a function zonal distribution of $\alpha$.

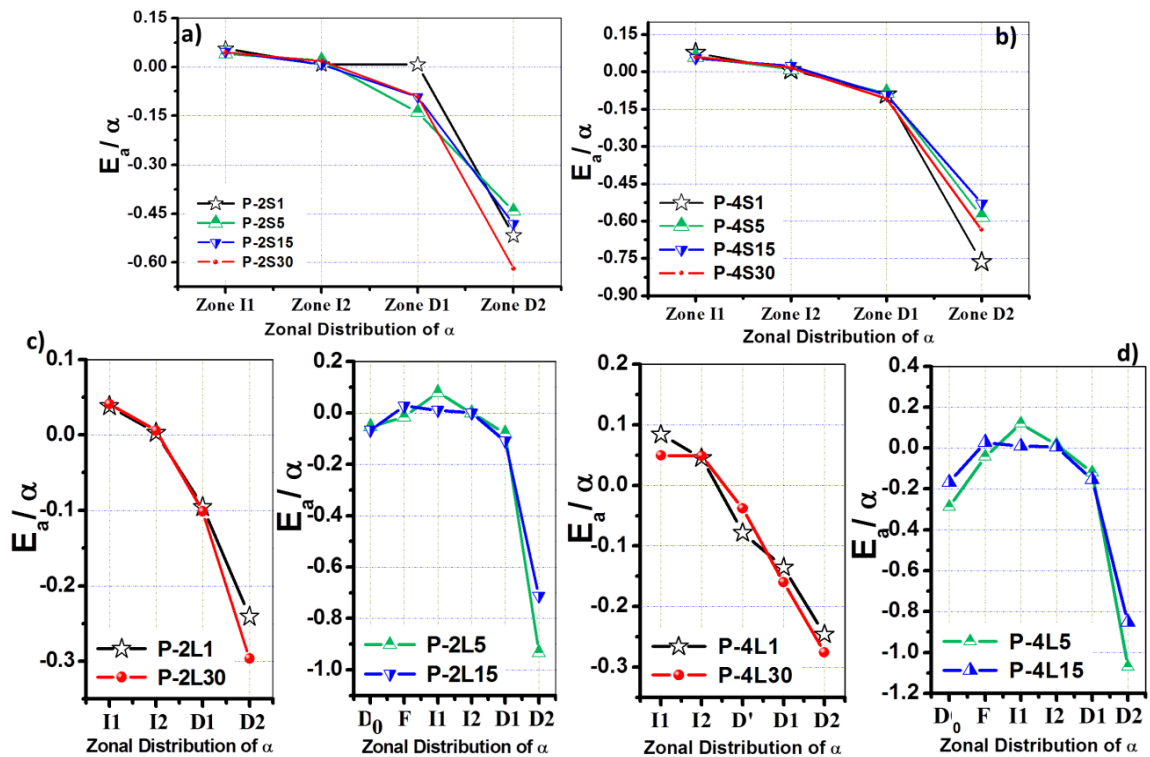

**Fig. 6.** Variation of Ea/$\alpha$ as a function of different zones of degree of conversion ($\alpha$) derived from the plot of Ea vs. $\alpha$ for: a) P-2S-series, b) P-4S-series, c) P-2L-series and d) P-4L-series.

Research articles have specified the melting process of PEO into following steps as: a) Onset of discernible melting (~ 52°C), b) Initiation of melting (~ 63°C), c) Partial melting (~ 64°C) and d) slow melting (completion of melting process) [~ 64-65°C] [29]. The early stages of melting comprises of sperullites development, more specifically described as *Hedritic*. Before the onset of crystallization stack of lamellae at the core of the *Hedrite* splay out, imparting an overall sphericity to the structure. This is indicated by the ascending portion of $E_a$ vs. $\alpha$ graph i.e Zone I. For P-2/4L-5 & 15, Zone $D_0$, F and I1 signifies such onset process. This is followed by initiation of melting which correspond to the featureless patches dispersed throughout the matrix. This continues till the presence of both longer dominant lamellae and shorter subsidiary lamellae. Along this there exist molten region as well. This is signified by *Zone I2*, upto which trend of $E_a$ is increasing as also depicted in the constant $E_a/\alpha$ change in Fig. 6. Lastly, the final process of slow melting initiates in which all traces of crystalline structure started disappearing. The activation barrier starts reducing and is exhibited by *Zone D', D1* and *D2*. For P-4L1 & 30, the slow melting process get initiated at ~ $\alpha > 40$ % which is also evident form the negative slope in Figs. 6c and d. Slow melting consists of a second stage reorganization process in which the PEO system shows negative activation energy, Zone D2. The system at this stage does not require external energy for completion of melting. Irrespective of the progression of melting process ($\alpha$) and variation in zonal divisions for all samples, the activation energy for P-2/4 S-series follows the trend as:

$E_a$ (P-2/4SUn) < $E_a$ (P-2/4S15) < $E_a$ (P-2/4S5) < $E_a$ (P-2/4S30) < $E_a$ (P-2/4S1)

Initiation of gamma irradiation of 1kGy enhances the activation barrier to the highest throughout the $\alpha$ regime. Subsequent increase in irradiation upto 15 kGy reduces the activation energy of the melting process till $\alpha = 70$ %. However, at the highest dose of 30 kGy, the energy requirement increases but remains lower than P-2/4S1. Compared to such erratic behaviour of $E_a$ in P-2/4S-series, P-2/4 L- is found to follow a regular trend as:

$E_a$ (P-2/4LUn) < $E_a$ (P-2/4L1) ≈ $E_a$ (P-2/4S30) < $E_a$ (P-2/4S15) < $E_a$ (P-2/4S5)

In P-2/4L-series, the activation energy chronologically increases till 5 kGy followed by a reduction till 30 kGy which is nearly equal to the $E_a$ of P-2/4L1. Highest crystallization of such P-2/4L-sreies in the region of 5-7 kGy (Fig. 1) tends to raise the activation of melting process. This is followed by a steady decrease in $E_a$, which signifies the formation of higher amorphous region compared to P-2/4 S-series.

### 3.3. Influence of irradiation dose and sample phase on properties of Polymer

*3.3.1. X-ray diffraction analysis*

The X-ray diffraction pattern of the synthesized PEO films obatined from irradiated powder and methanol solution is shown in Figs. 7 and 8.

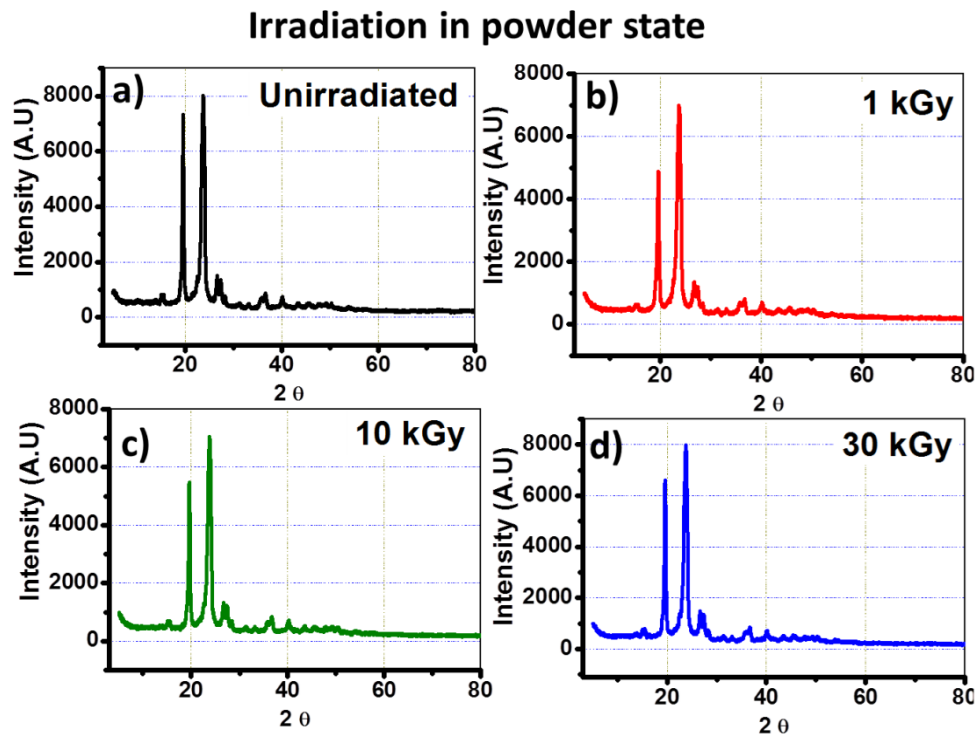

**Fig. 7.** X-ray diffraction pattern for a) unirradiated PEO film and film obatined from powder irradiation at doses of: b) 1kGy, c) 10 kGy and d) 30 kGy.

Fig. 7a shows the XRD pattern of PEO film without any irradiation history. The indexing of PEO peaks are reported to be (120) at 19.5° and (032) + (112) at 23.5° of the monoclinic system [30]. Both Figs. 7 and 8 exhibit the formation of two sharp well defined peaks which signifies pure polymer system with crystalline region in significant proportion.

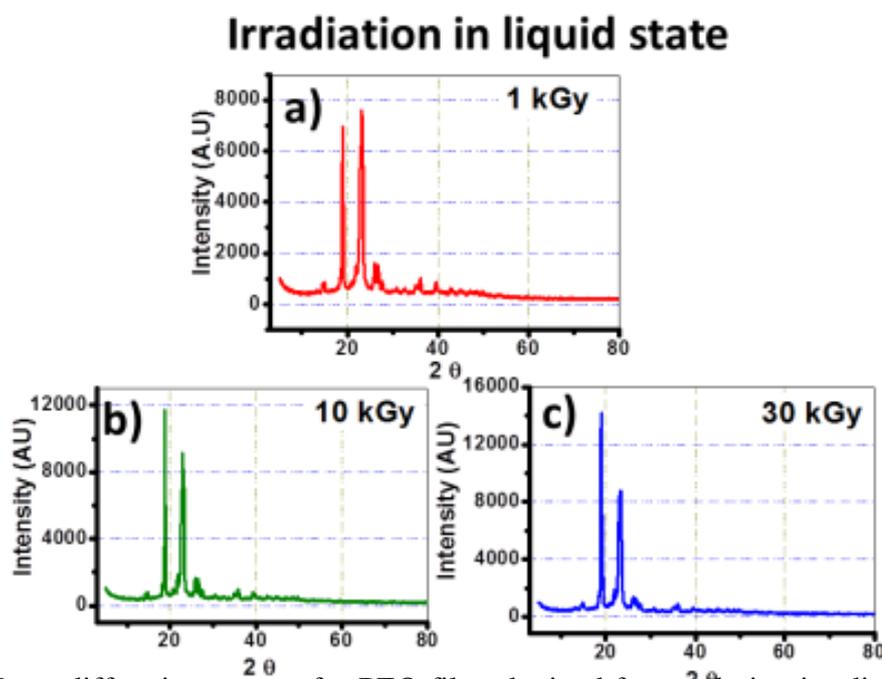

**Fig. 8.** X-ray diffraction pattern for PEO film obatined from solution irradiation at doses of: b) 1kGy, c) 10 kGy and d) 30 kGy.

The peak positions of unirradiated PEO films and that obatined from powder irradiation appear at $19.56^o$ and $23.7^o$ (Fig. 7), wheareas in case of films obatined from solution irradiation, the same is observed at $18.9^o$ and $23.3^o$ respectively. The higher intensity of XRD peaks for PEO films obtained from solution irradiation clearly signifies greater crystallinity which is also supported from the DSC study (Fig. 1). In Fig. 7, a lowering of the first peak is observed with increase in irradiation dose, whereas, irradiation in solution state (Fig. 8) lowers the second peak with similar increase in dose. In addition, the extent of lowering of either peak is almost independent of dose rate in Fig. 7, whereas, in Fig. 8, the degree of second peak depression is more pronounced at higher dose. This could be explained on the fact that, in case of solution irradiation, the % crystallinity is found to increase upto 7 kGy followed by a steady reduction (Fig. 1). In spite of the higher crystallinity of PEO films of P-2L or P-4L –series upto 5 kGy, the same for higher doses is significantly lower compared to even P-2S or P-4S-series.

*3.3.2. Structural modifications and prediction of reaction processes: A study using FTIR*

The absorptions in the infrared spectra of the experimental PEO samples (both unirradiated and irradiated) are assigned according to the generally accepted convention by Yosihara [31]. Gamma degradation process involves initiation, propagation, and termination stages. The initiation reaction may take place at different sites of the PP chain. When energy is absorbed from gamma rays, it causes degradation of the covalent backbone. The isotactic arrangements as well as the -C-H bond of the methylene group breaks followed by the formation of many newer products.

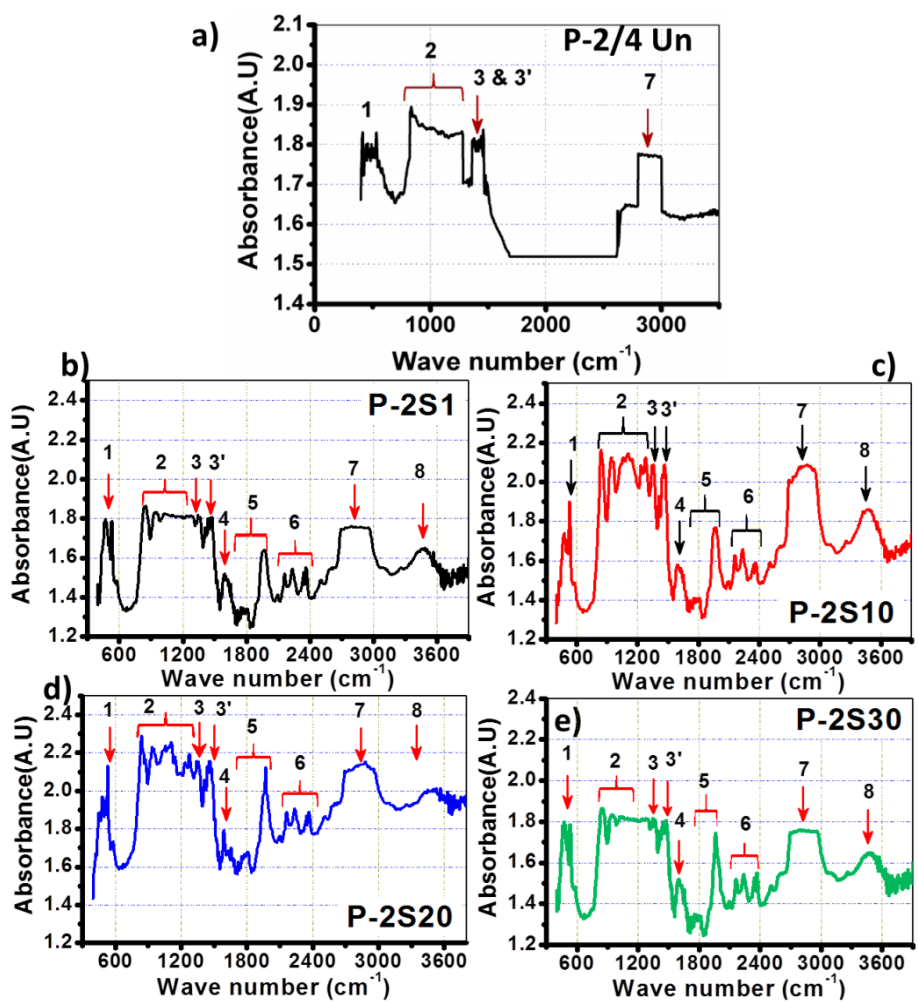

**Fig. 9.** FTIR spectra for: a) unirradiated PEO and film obtained from powder irradiation at doses of: b) 1kGy (P-2S1), c) 10 kGy (P-2S10), d) 20 kGy (P-2S20) and e) 30 kGy (P-2S30).

Figs. 9 -12 shows the corresponding FTIR results of PEO films obtained from both irradiated powders and solution of variable concentration. The characteristic identification of the FTIR peaks indicated by identification numbers is illustrated in Table 2 along with the absorption frequencies.

**Table 2:**
Sample identification of experimental PEO samples subjected to gamma irradiation

| Peak ID | Wave Number (cm$^{-1}$) | Interpretation |
|---|---|---|
| 1 | 460-530 | Finger print region, bending vibrations |
| 2 | 850-1270 | -C-O-C- stretching vibration |
| 3 and 3' | 1350-1450 | -CH$_2$- bending vibration (waging) |
| 4 | 1590-1650 | frequency of -C= C- |
| 5 | ~1730-1750 | Shows the presence of -C= O group which may correspond to the absorption of shorter wave length by carboxyl, ketone, aldehyde groups |
|   | 1960 | -C= O stretching modes of a dimer molecule |
| 6 | 2148-2360 | Frequency of 'O≡C'→M terminal mode |
| 7 | 2800-3000 | Methylene stretching (-C-H-) |
| 8 | 3300-3500 | Stretching frequency of -OH group |

With reference to Fig. 1, FTIR spectra for only 1, 10, 20 and 30 kGy are given for P-2/4S-series and 1, 7, 20 and 30 kGy are shown for P-2/4L-series. FTIR spectra of the unirradiated PEO film (P-2/4 Un) [Fig. 9a] shows normal frequencies corresponding to usual bending vibrations in finger print region (Peak 1), stretching vibrations of –C-O-C (Peak 2) and methylene stretching frequencies (Peak 7). The doublet –CH$_2$- bending vibrations in the range of 1350-1450 cm$^{-1}$ (Peak 3 and 3') indicates the amorphous nature of PEO. Absence of sharp peaks in this region could be explained due to lesser crystallinity % in the matrix (also supported from Fig. 1). Perturbation of PEO system either in the powder or solution state by gamma irradiation is found to generate many additional peaks which prove the formation of newer products.

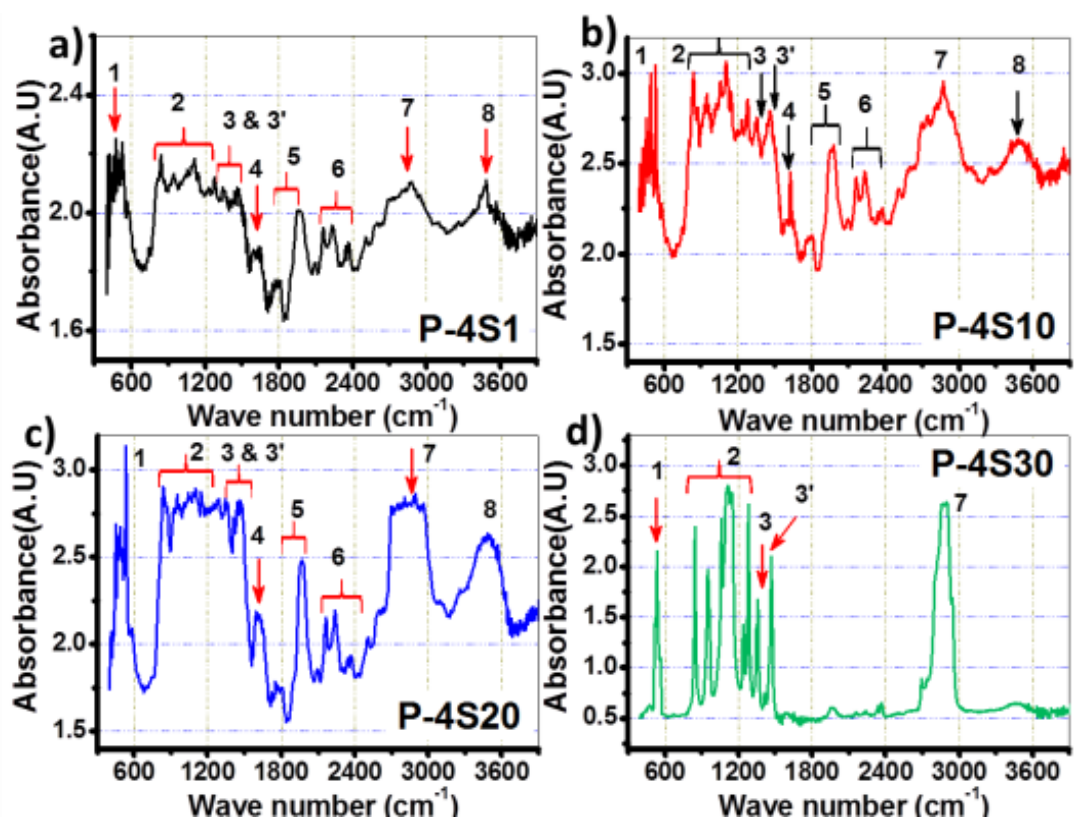

**Fig.10.** FTIR spectra for PEO film obtained from powder irradiation at doses of: a) 1kGy (P-4S1), b) 10 kGy (P-4S10), c) 20 kGy (P-4S20) and d) 30 kGy (P-4S30).

In P-2S or P-4S system, sharp peaks of –$CH_2$- bending vibrations (Peak 3 and 3') are formed compared to Fig. 9a. These signify the increment in crystallization within the matrix. In addition, the trend of % crystallinity obtained from DSC studies (Fig. 1) for P-2/4S-series shows similar order of crystallinity enhancement upto 20 kGy which could be associated with the magnitude of intensity of the FTIR plot. This is followed by a sharp reduction in crystallinity at 30 kGy. Similarly, the order of % crystallinity as a function of irradiation dose is satisfied by P-2L-and P-4L-series also (Figs. 11 and 12). The stretching vibration of -C-O-C- bond (Peak 2) exhibits multiple overlapping peaks which are sensitive to the chain conformations that exist in the semi-crystalline PEO. Compared to powder state, the irradiation effect are more pronounced even at lower doses for liquid state due to greater mobility of polymer chains and easy diffusion of intermediate short lived radicals [32]. As expected -C-O-C- stretching peaks are more sharp and intense for P-2/4L-series. In liquid irradiation, the contribution of cross linkage is more due to the formation of –OH radicals from solvent which acts as the cross linking agents.

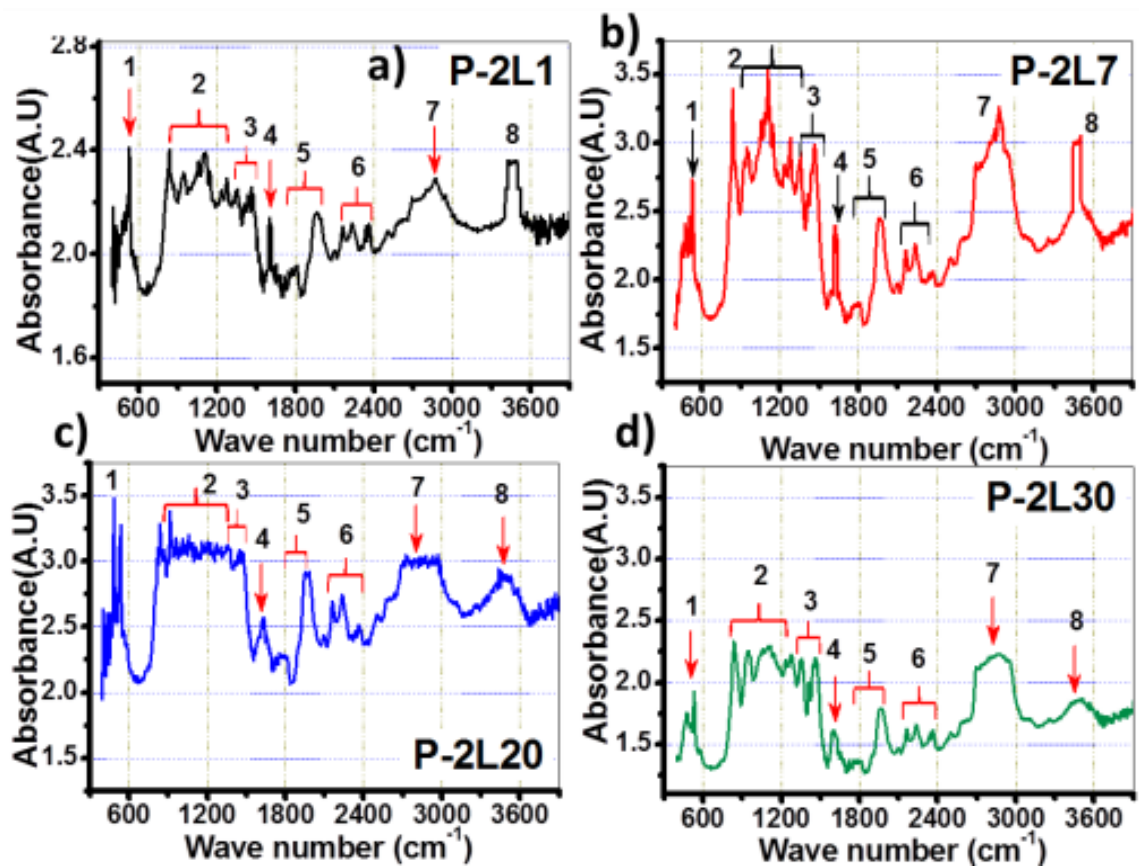

**Fig. 11.** FTIR spectra for PEO film obtained from liquid irradiation at doses of: a) 1kGy (P-2L1), b) 7 kGy (P-2L7), c) 20 kGy (P-2L20) and d) 30 kGy (P-2L30).

In spite of the scission dominance due to air induced irradiation, PEO samples of P-2/4L-series shows enhanced cross linkage compared to P-2/4S-series (also evident from Fig. 1, where crystallinity starts reducing after 7 kGy for P-2/4L-series). This is also supported from the widening of Peak 2 in Figs. 11 and 12. Significantly higher intensity of O-H stretching frequency (Peak 8) for P-2/4L-series further proves the generation of such cross-linking agents upon liquid state irradiation. Presence of –O-H group furthers supports the formation of hydrogen bonding which is higher in P-2/4L –series. A signature of degradation/scission upon irradiation is observed from the formation of -C= C- and -C= O group (Peak 4-6).

Irrespective of sample configurations, it could be observed that at the lower doses, the contribution of Peak-4 (-C= C- stretching) is significant but with dose increment the concerned intensity reduces compared to Peak 5 (-C= O group) which increases significantly.

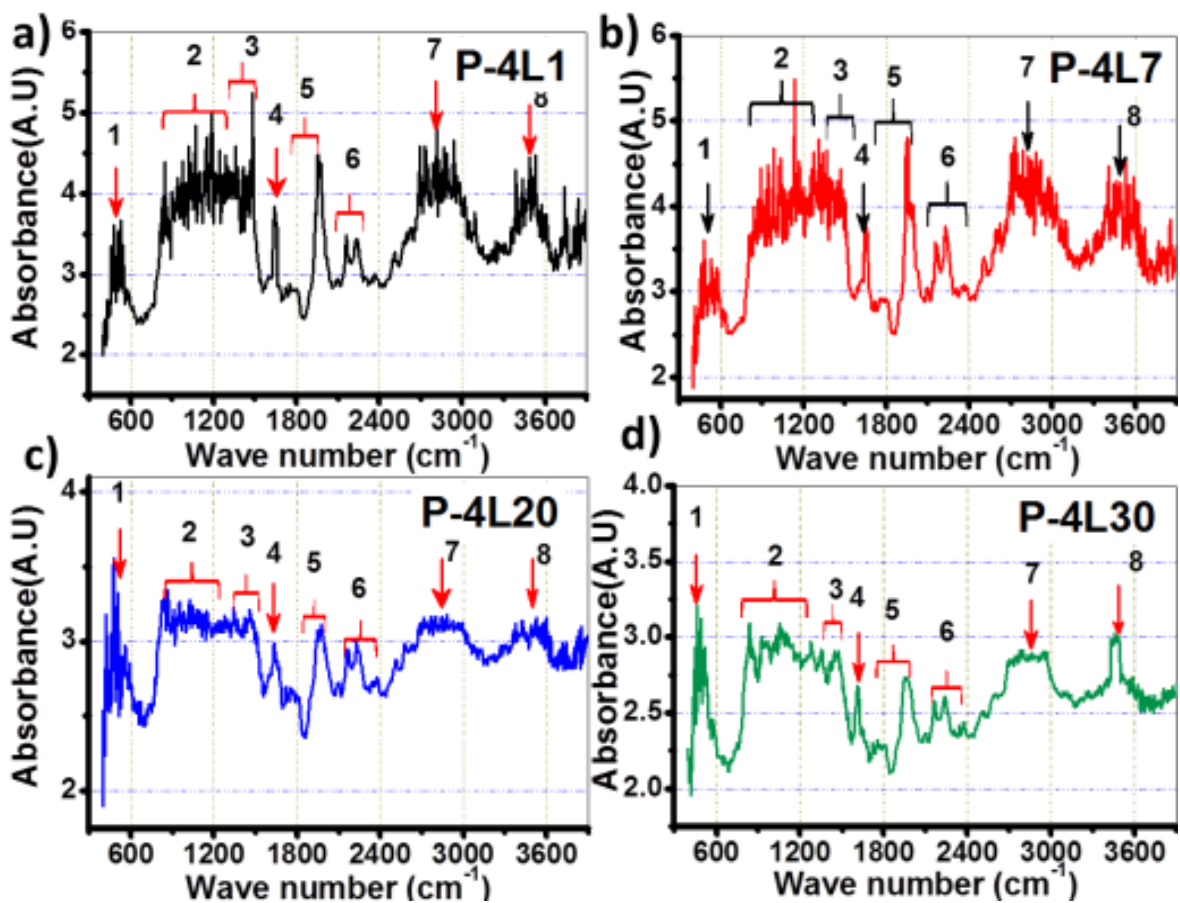

**Fig.12.** FTIR spectra for PEO film obtained from liquid irradiation at doses of: a) 1kGy (P-4L1), b) 10 kGy (P-4S10), c) 20 kGy (P-4L20) and d) 30 kGy (P-4S30).

This could be explained on the fact that -C= C- bonds form as a result of degradation and disproportionation in regions which lack oxidation degradation due to lower concentration of $O_2$. In the present air driven irradiation of PEO, concentration of oxygen increases with dose thereby increasing the intensity of Peak 5. Higher contribution of oxygen induced degradation is further supported by the presence of Peak 6 which proves the formation of terminal carbonyl groups.

A pronounced peak at ~ 1960 cm$^{-1}$ (part of Peak 5) is observed in all samples which shows the signature of -C= O group of a dimer molecule. Such formation proves the simultaneous occurrence of cross linkage within PEO matrix.

FTIR spectra of P-4S-30 is entirely different from other irradiated samples, whereas it bears similarity with P-2/4 Un. All the peaks corresponding to functional groups viz. 4, 5, 6 and 8 are absent and only signify the present of simple PEO entity. This could be explained by the fact that, highest irradiation of 30 kGy onto concentrated PEO matrix tend to disintegrate the spherullite to such an extent at that its size could not be reported. This is shown in the polarizable microstructure shown in the inset of Figs. 13c2-I1. From Fig. 1, it is evident that the % crystallization of P-4S30 is not significantly lowered compared to P-2/4L. Whereas, the extent of degradation is maximum with negligible contribution of cross-linkage for P-4S30. Though, numerous short chains are formed which acts as the nuclei site for crystallization, but the reaction proceeds in such a way so as to eradicate the formed functional groups and therefore, only representative FTIR bands of PEO chain is evident.

*3.3.3. Correlation with Microstructural analysis*

SEM micrographs of PEO films obatined from powder and solution irradiation with variable concentration is shown in Figs. 13 and 14 along with the image of unirradiated PEO film. Compared to the morphology of the unirradiated PEO film, irradiation generates strands or fibrils within the PEO matrix. Though the unirradiated PEO matrix possess crystallinity, the SEM images does not exhibit the formation of such fibrils and rather show homogenous region at high magnification (Fig. 13 a2). Owing to such fact, a low magnification image (Fig. 13 a1) is shown for unirradiated matrix. It could be noted that in P-S-series (Fig. 13b-e), with increase in irradiation dose upto 20 kGy, the recognized fibrils or small branches are numerous. These fibrils are absent in the morphology of PEO irradiated for highest 30 kGy. Formation or presence of fibrils signifies the presence of crystalline regions which increases upto 20 kGy for P-S-series (Fig. 13) and 7 kGy for P-L-series (Fig. 14). Air induced degradation at highest dose of 30 kGy onto PEO powder disintegrates the polymer strands (already discussed in section 3.3.2).

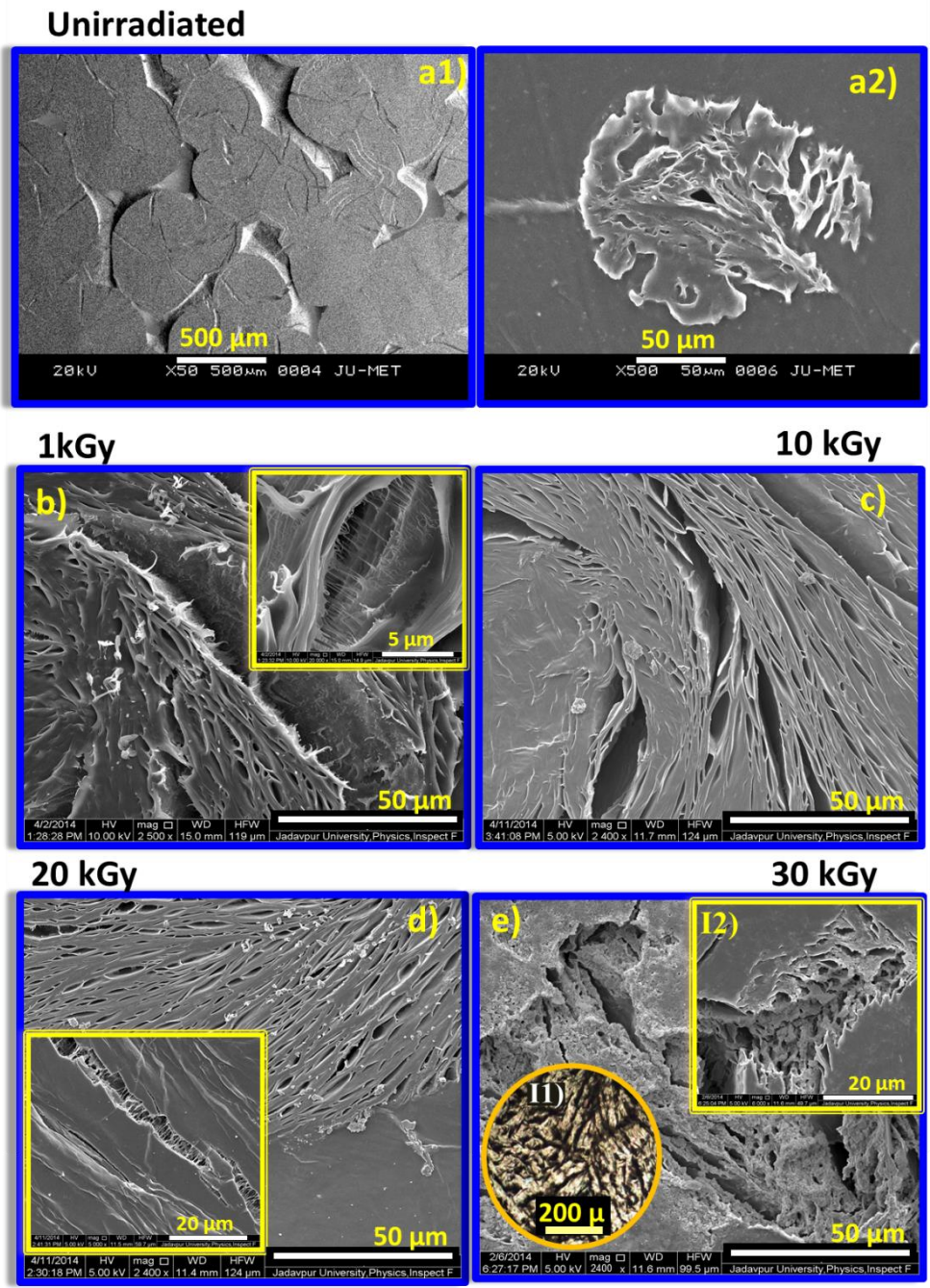

**Fig. 13.** SEM micrograph for: a1), a2) unirradiated PEO film and film obtained from powder irradiated at doses of: b) 1kGy, c) 10 kGy, d) 20 kGy and e) 30 kGy. Inset: Images of b), d) and e-I2) are the higher magnified SEM images and e-I1) polarizable microstructure of PEO film from powder irradiated at 30 kGy.

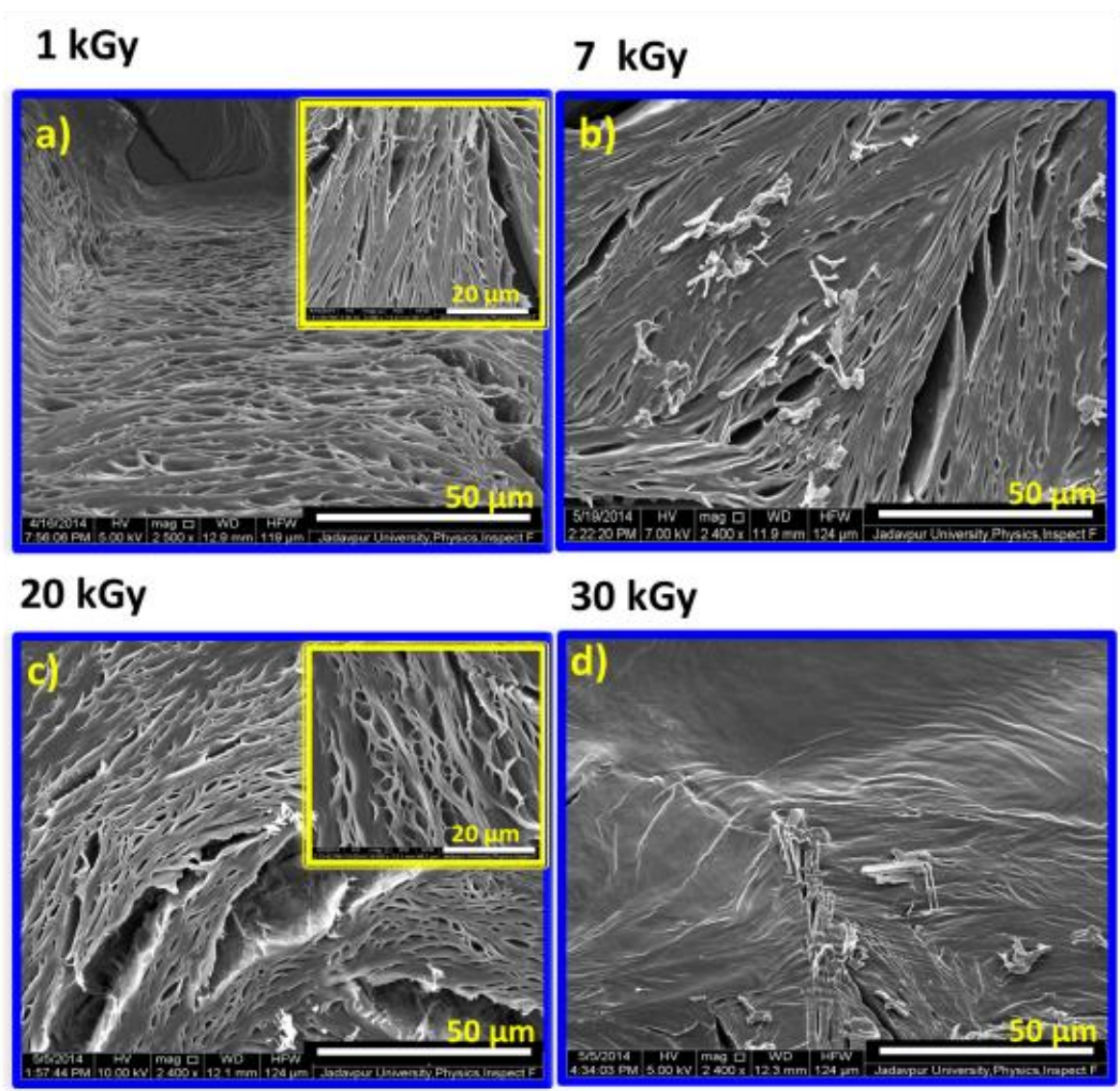

**Fig. 14.** SEM micrograph of PEO film obtained from liquid irradiated at doses of: a) 1kGy, b) 7 kGy, c) 20 kGy and d) 30 kGy. Inset: Images of a) and c) are the higher magnified SEM images.

Such totally disintegrated matrix is shown in the inset of Figs. 13 e-I1. Higher % crystallinity of the PEO film obatined from solution irradiation (Fig. 1) is also supported from Figs. 14 a and b which shows much higher formation of polymer fibrils compared to the films of P-S series. Stojanovic´ et al., Zhang et al.etc recognized the presence of tie molecules/ fragments in the interfacial regions that connects the crystallites [19, 20]. They proposed recrystallization of such tie entities as another cause of the increase in crystallinity at lower doses. Formation of such tie entities could be visualized from the inset of Fig. 13b. These tie molecules tend to disintegrate with increase in dose (inset of Fig. 13c) and vanishes at 30 kGy dose (inset of Fig. 13d). The tie molecules are also observed in the PEO film obtained from liquid irradiation. Though the % crystallinity of P-2/4L series at 30 kGy is much lower than P-2/4S, but the SEM micrograph (Fig. 13d and Fig.14d) corresponding to the mentioned batches vary significantly. The matrix of P-2/4L30 clearly reveals the presence of more cross-linking agent generated from solvent upon irradiation compared to P-2/4S30. Therefore, selected irradiations could be utilized in a specific manner so as to get optimized results in terms of crystallinity, generation of functioned groups etc. according to the functional requirement.

4. **Conclusions**

Combined DSC and FTIR study have revealed that the influence of perturbation on Poly [Ethylene Oxide] (PEO) by gamma irradiation varies significantly with the physical state [solid or solution]. The irradiated (1-30 kGy) PEO powders (P-S-series) as well as solutions (P-L-series) are casted into films having 2 g.ml$^{-1}$ and 4 g.ml$^{-1}$ concentration and are studied for property evaluation. DSC results exhibit steady increment of crystallinity upto 20 kGy for P-S-series after which amorphous region increases till 30 kGy. Conversely, P-L-series shows much enhanced crystallinity retained within low regime of 7 kGy, followed by sharp declining trend till 30 kGy. These observations are also supported from the trend of crystallization enthalpy ($\Delta H_c$) determined from the cooling cycles. Multiple melting and cooling cycles are carried out in order to establish the above observations upon erasing the previous history. DSC is also used to determine the multiple kinetic processes as a function of irradiation dose for P-2/4S and P-2/4L -series in an isoconversional PEO melting using Friedman differential analysis. The experimental samples show variable dependence of activation energies ($E_a$) as a function of degree of melting conversion ($\alpha$) which could be confined into various zones. The increasing and decreasing trend of $E_a$ as a function of $\alpha$ could be divided into four zones for P-2/4S-series and P-2L-1 & 30, whereas, for P-4L-1 & 30, an additional *Zone D′* could be identified which exhibit early decreasing of $E_a$ from $\alpha \approx$ 45 %. Compared to all samples, P-2/4L-1 & 15-batches initially consumes maximum energy for small $\alpha$ (observed from the highest $E_a$ during starting of the melting), then proceeds without barrier upto $\alpha = 20$ % (Flat region: *Zone F*). This is followed by ascending of activation barrier and finally decreases at higher $\alpha$. The negative activation energy for all samples at $\alpha > 90$ % signifies self-sustaining reaction/ process which does not require external activation, whereas, the internal energy is sufficient to complete the isoconversional process. The various kinetic regions thus obatined are correlated with the melting stages viz. discernible melting, initial, partial and finally slow melting. Irradiation is found to generate various new functional groups as studied from FTIR study. Compared to powder state, the irradiation effect is more pronounced even at lower doses for liquid state due to greater mobility of polymer chains and easy diffusion of intermediate short lived radicals. Formation of -C=C- and -C=O group proves the dominance of degradation in air induced irradiation process. With increase in dose, intensity of -C=O increases due to higher availability of $O_2$ from air. Higher contribution of oxygen induced degradation is further supported by the formation of terminal carbonyl groups. Contribution of cross-linkage is more in films formed by liquid irradiation due to the formation of –OH radicals from solvent which acts as the cross linking agents and is proved from higher intense peak of –OH group. This reduces crystallinity after 7 kGy which is supported form widening of –C-O-C- stretching vibration. In addition, a low intense peak of –C=O dimer is observed which further supports the co-existence of cross-linkage along with scission/ degradation. These mentioned observations are also supported by SEM micrographs. The micrographs clearly exhibit the formation tie linkages between polymer fibrils/ strands which signify the crystalline zones.

With increase in irradiation beyond 20 kGy for P-S-series and after 7 kGy for P-L-series, tie linkages tend to break thereby increasing the amorphous regions in the polymer matrix. Hence, selective irradiation dose could be determined with respect to the exposed state (solid or solution) of polymer and utilized to tailor the properties of PEO.

## Acknowledgements

The research work is funded by UGC Major Research Project [F. No. 41-847/2012 (SR)] for financial support and MM is thankful to Council of Scientific and Industrial Research (CSIR), India for providing Research Associate fellowship. Dr. Paramita Bhattacharyya and Miss. Suchisrawa Ghosh of Department of Food Technology, Jadavpur University, Kolkata are acknowledged in extending their support in using the gamma irradiation chamber. The Authors also acknowledge FIST-2, DST Government of India, at the Physics Department, Jadavpur University for providing the facility of SEM microscope.